\documentstyle[psfig,conf_iap,10pt]{article}

\def\etal{{\it et al.\ }}

\begin{document}

\heading{The Compared Number Density of High-Redshift \\ Galaxies and Lyman 
$\alpha$ Clouds} 

\par\medskip\noindent

\author{A. Fern\'andez-Soto$^{1,2}$, K.M. Lanzetta$^{2}$, A. Yahil$^{2}$, 
H.-W. Chen$^{2}$}

\address{Dept.\ of Astrophysics and Optics, UNSW, Sydney, NSW 2052, AUSTRALIA}
\address{Dept.\ of Physics and Astronomy, SUNY, Stony Brook, NY 11794-3800, 
USA}

\begin{abstract}
 We use our catalog of photometric redshifts in the Hubble Deep Field (HDF) 
to estimate the Luminosity Function (LF) of galaxies up to $z=2$. Using the 
obtained LF and a relationship between luminosity and halo size, we calculate 
the expected density of galactic halo crossings for any arbitrary line of 
sight. This density is then compared with the known one of Lyman $\alpha$ 
lines, showing that the observed density of galaxies is enough to account for 
the observed absorption lines. 
\end{abstract}

\section{Galaxies, Luminosity Functions and QSO Absorbers}

 We have obtained a photometric redshift catalog \cite{fs97} of objects in 
the HDF \cite{wil96}. The catalog is complete to $AB(8140) = 28.0$, and it has 
been measured to be accurate to $\Delta z_{rms}=0.15$ when compared to 
available spectroscopic values.  As the HDF observations cover a wide 
wavelength range (3000--8000 \AA), we can obtain the $B$-band rest-frame flux 
of every object with $z<1.0$ and the $U$-band rest-frame flux of those with 
$z<2.0$ without the need of applying any K-correction. This advantage, 
together with the unprecedented depth of the HDF images, allows us to 
estimate the $B$- and $U$-band LF down to very faint absolute magnitudes, and 
to break it in several redshift bins. Complete details of this procedure will 
be presented elsewhere, but we  mention here that the obtained LFs show a 
steep increase in number in the faint end (at $M>-15$) and that there is no 
evidence for strong evolution up to $z=1$, although number evolution might be 
necessary to fit the measured LF at higher $z$.

 Recently Chen \etal \cite{hw97} have shown the existence of a relationship 
between the luminosity of a galaxy and the size of the gaseous halo around 
it: $\rho \propto (L_B/L_B^*)^{0.31}$, where $\rho$ is the minimum impact 
parameter at which a galaxy with luminosity $L_B$ will produce an absorption 
line with EW$> 0.3$ \AA\ (using $H_0=100\ {\rm km\ s^{-1}\ Mpc^{-1}}$ and 
$q_0=0.5$, \cite{hw97} obtain $\rho=120\ {\rm kpc}$ for a $L^*$ galaxy). This 
relationship, used in conjunction with the obtained LFs, allows us to 
calculate $dN_{HC}/dz$, the average expected density of ``halo crossings'' 
per unit redshift for a line of sight to a background QSO that would produce 
such lines. These numbers are plotted in Figure 1 (left), with different 
marks corresponding to the values obtained from the $B$- and $U$-band LFs 
at different redshifts. We have integrated the functions only down to 
$L=0.01L^*$, although in none of the actual cases the function diverges when 
$L$ tends to zero.

\section{Comparison with QSO Absorption Lines}

 The results described in \cite{hw97} refer to galaxies over the redshift 
range 0.1--1.2. We are extending those results towards slightly higher 
values of $z$, but as high-$z$ galaxies are expected to have even higher gas 
contents than their low-$z$ counterparts, this represents a conservative 
estimate.

 A direct comparison can be established with the known redshift evolution of 
Lyman $\alpha$ absorbers, $dN_{Ly\alpha}/dz$ (\cite{bah}, \cite{bech}; see 
Figure 1 right). The plots show clearly that galaxies alone are able to 
account for {\em all} the absorption lines produced in QSO spectra up to 
redshifts $z \approx 2$. There is also a hint of a break in the function 
$dN_{HC}/dz$ that can be compared to the one in $dN_{Ly\alpha}/dz$, although 
this result must be tested with LFs that extend to higher redshift. 

 These results, when taken together with recent measurements of the metal 
abundances of Lyman $\alpha$ clouds \cite{cow95}, their clustering 
properties \cite{fs96} and the optical identification of the galaxies 
responsible for Lyman $\alpha$ absorption on QSOs \cite{ken95}, lead us to 
maintain that there seems to be little or no place for the hypothesis of a 
second intergalactic population of Lyman $\alpha$ clouds.

\begin{figure}
\centerline{\hbox{
\psfig{figure=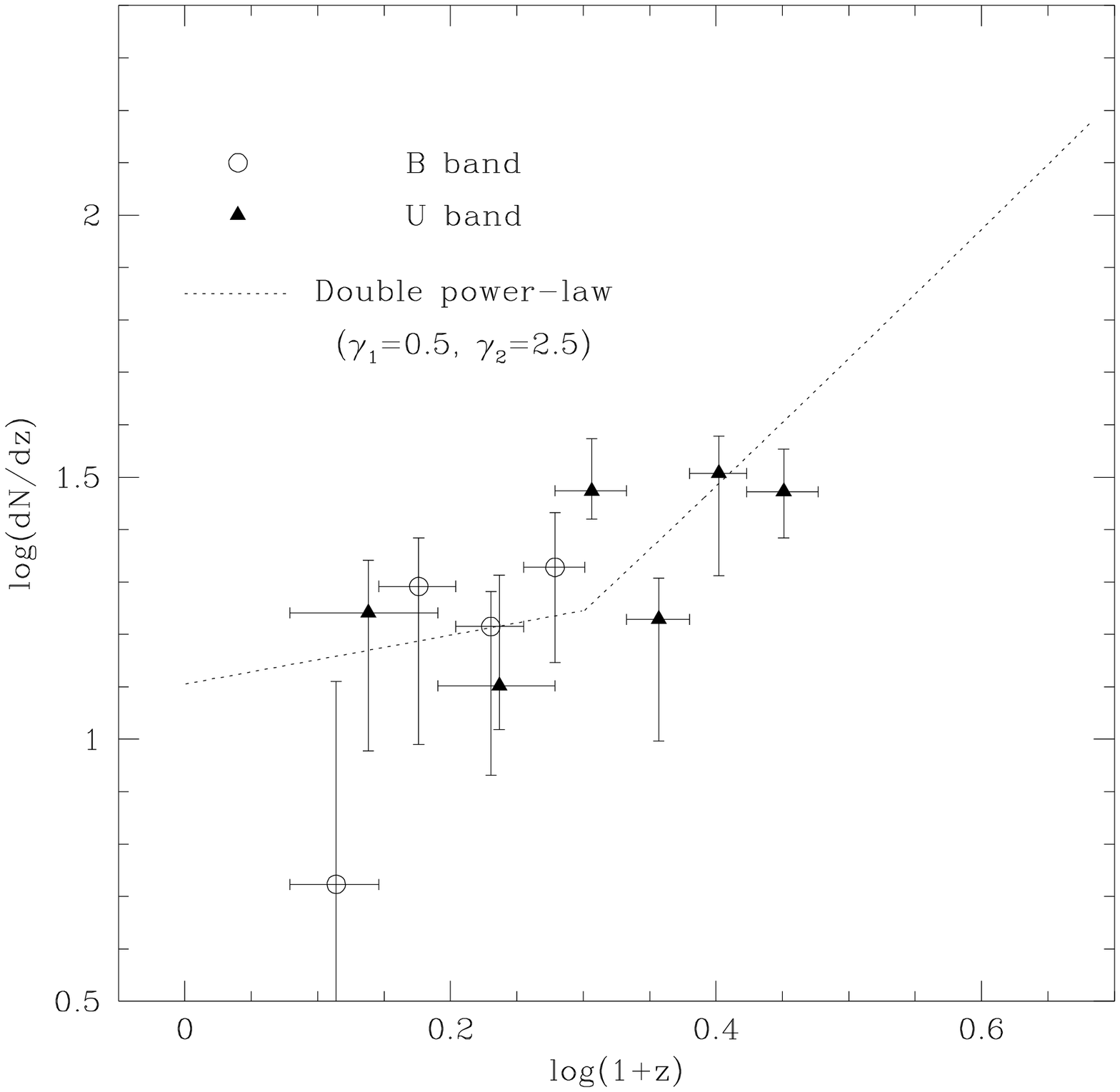,height=4.9cm,width=6cm}
\psfig{figure=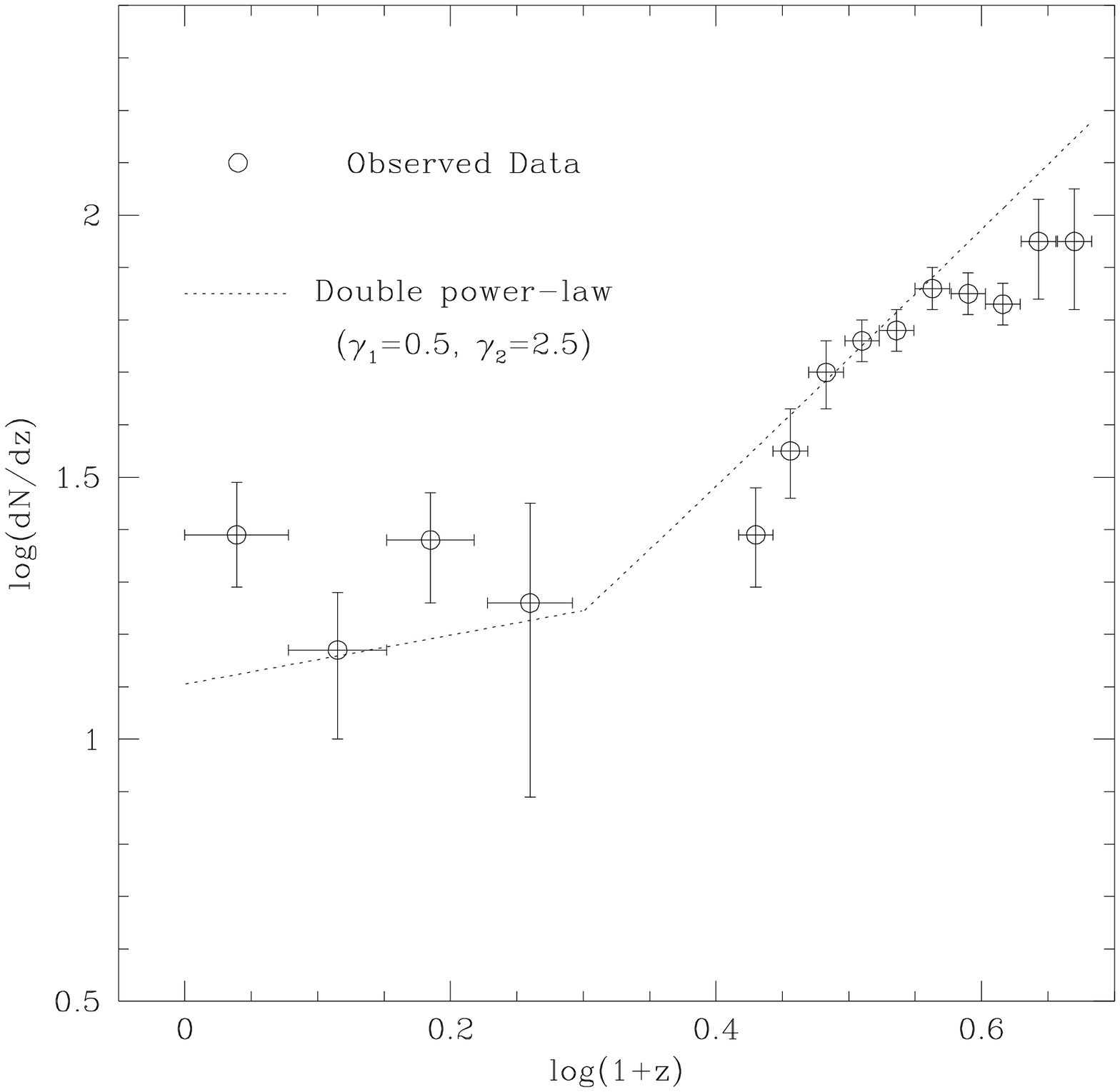,height=4.9cm,width=6cm}
}}
\caption[]{Redshift density of ``halo-crossings'' (left) and Lyman $\alpha$ 
absorbers (right). The dotted line is a typical double power-law fit used only 
as a guide.}
\end{figure}



\vfill
\end{document}